\documentstyle[twocolumn,epsf]{jpsj}

\def\runtitle{Quasiparticle Properties around a Nonmagnetic Impurity}
\def\runauthor{Hiroki {\sc Tsuchiura}, Yukio {\sc Tanaka},
 Masao {\sc Ogata} and Satoshi {\sc Kashiwaya}
}

\title
{
Quasiparticle Properties around a Nonmagnetic Impurity 
in \\
the Superconducting State of the Two-Dimensional $t$-$J$ Model
}
\author
{ 
Hiroki {\sc Tsuchiura}$^{1}$, Yukio {\sc Tanaka}$^{2,1}$,
Masao {\sc Ogata}$^{3}$ and Satoshi {\sc Kashiwaya}$^{4,1}$
}

\inst
{
$^{1}$CREST, Japan Science and Technology Corporation (JST), Nagoya, 464-8603\\
$^{2}$ Department of Applied Physics, Nagoya University, Nagoya, 464-8603\\
$^{3}$ Department of Basic Science, Graduate School of Arts and Sciences,
University of Tokyo,\\
 Komaba, Meguro-ku, Tokyo, 153-8902\\
 $^{4}$ Electrotechnical Laboratry, Umezono, Tsukuba, Ibaraki, 305-8568\\
}

\recdate
{
}

\abst
{
The superconducting state around a well-isolated nonmagnetic impurity in the high-$T_{c}$ superconductors is studied using the two-dimensional $t$-$J$ model.
The spatial dependence of the order parameter and the local density of states are obtained from the numerical diagonalization of the Bogoliubov-de Gennes equation derived using the Gutzwiller approximation.
We find a zero-energy peak in the local density of states around the impurity.
Different from the previous results on a vortex or surfaces in the $t$-$J$ model, the splitting of the zero-energy peak is not found.
The zero-energy states corresponding to the zero-energy peak is approximately localized around the impurity.
}

\kword
{
Two-dimensional $t$-$J$ model, Gutzwiller approximation, $d$-wave superconductivity, a nonmagnetic impurity, zero-energy peak
}

\begin{document}
\sloppy
\maketitle

The role of impurities in superconductors has been studied theoretically
since the establishment of BCS theory.
In the conventional ($s$-wave) superconductors, nonmagnetic impurities
have only little effect on the transition temperature, as understood
from Anderson's theorem \cite{pw}.
On the other hand, they have strong effects on the superconducting
properties of unconventional (anisotropic) superconductors in various ways depending on the anisotropy of the pairing state.
There have been many studies on the nonmagnetic impurity effects in connection with heavy fermion\cite{hf1, hf2, hf3} and high-$T_{c}$ superconductors\cite{hotta, lee, hirschfeld, sun, onishi, bala, bala2, flatte, franz}.

Regarding high-$T_{c}$ superconductors, a large number of experimental results and theories support the $d_{x^2-y^2}$-wave pairing state.
The significant difference between the $d_{x^2-y^2}$-wave pairing state and the conventional one is that, in the $d_{x^2-y^2}$-wave state, the order parameter (pair potential) changes its sign with $\pi/2$ rotation.
Thus, the nodes of the energy gap exist on the Fermi surface along the $k_{x} = \pm k_{y}$ directions in the Brillouin zone.
The existence of the nodes implies that there are residual quasiparticles around the nodes.
Thus, the $d_{x^{2}-y^{2}}$-wave pairing state has {\it an intrinsic instability} \cite{ogata,laughlin}.
This instability could show up in nonuniform systems, such as those with vortices, surfaces or impurities.
In these cases, we expect an interference effect of the quasiparticles because the quasiparticles feel the sign change of the pair potential when they are reflected.
Actually this interference effect causes the zero-energy states, which can be detected as the zero-energy peak (ZEP) in the scanning tunneling spectroscopy near the vortex core \cite{maggio} and surfaces \cite{alff}.

In this paper we study this interference effect in the nonmagnetic impurity problem.
To date, there have been some works on the nonmagnetic impurity in the $d_{x^{2}-y^{2}}$-wave superconductivity.
It has been shown that the lowest eigenvalue approaches zero as the impurity scattering approaches the unitary limit \cite{onishi,bala}.
This is the zero-energy state which is located around the impurity.
In this paper, we show for the first time that this state actually gives the ZEP, which can be observed in scanning tunneling spectroscopy around impurities.
In order to obtain the local density of states around a well-isolated impurity, we study fairly large systems.
Furthermore, we will show that the zero-energy state is localized around the impurity.
Our result gives an answer to the inconsistency between the previous theories, that is, whether the localized state has a slow decay ($\sim 1/r$) along the nodes of the gap \cite{onishi,bala2} or not \cite{lee,franz}.

Finally, in contrast to the vortex cores\cite{hime} and surfaces\cite{tanuma}, we show that the extended $s$-wave component is not induced around the impurity, and thus there is no splitting of the ZEP.

As a model for high-$Tc$ superconductors, we use the two-dimensional
$t$-$J$ model.
Quasiparticle states near a nonmagnetic impurity in the $d_{x^{2}-y^{2}}$-wave superconductivity is studied for the case of a unitary scattering limit, which is discussed in the previous works and is relevant to the Zn-doped case.
The spatial variation of the superconducting order parameter around the impurity is determined self-consistently.

The Hamiltonian of the $t$-$J$ model with a nonmagnetic impurity is written as
\begin{eqnarray}
{\cal H} &=& -t\sum_{\langle i,j \rangle \sigma}
( c^{\dag}_{i\sigma} c_{j\sigma} + c^{\dag}_{j\sigma} c_{i\sigma} )
 - \mu\sum_{i,\sigma}c^{\dag}_{i\sigma} c_{i\sigma}
\nonumber \\
& & + \sum_{\langle i,j\rangle}
(J\mbox{\boldmath $S_{i}\cdot S_{j}$} - \frac{J_{N}}{4}n_{i}n_{j} )
+ \sum_{i\sigma}V_{i}^{\rm{imp}}n_{i\sigma} ~,
\end{eqnarray}
where $\langle i,j\rangle$ means the summation over nearest-neighbor pairs, and the impurity potentials $V_{i}^{\rm{imp}}$ are nonzero only on the impurity sites.
Hereafter, we use $t$ as an energy unit.
Since the constraint of no double occupancy is imposed on this Hamiltonian, 
it is difficult to carry out analytical calculations even in mean-field theories.
Here we consider $T=0$ variational theory.
The variational wave function is $P_{G}|\rm{BCS}(\Delta_{ij})\rangle $, where $P_{G}$ is the Gutzwiller projection excluding the double occupancy, $P_{G} = \Pi_{i} (1-n_{i\uparrow }n_{i\downarrow })$, and $|\rm{BCS}(\Delta_{ij})\rangle $ is the BCS meanfield solution with the site-dependent order parameter $\Delta_{ij} = \langle c^{\dag}_{i\uparrow}c^{\dag}_{j\downarrow }\rangle$ ($ij$ nearest-neighbor sites). 
In order to evalute the variational energy with the Gutzwiller projection, we use a Gutzwiller approximation\cite{Zhang,yoko}, in which the constraint is taken into account as a statistical average.
This approximation has been extensively studied in connection with the
variational Monte Carlo (VMC) calculation \cite{yoko}.
For the two-dimensional $t$-$J$ model, it has been shown that
the Gutzwiller approximation and the VMC calculation give very
similar variational energies \cite{yoko}.

Using the Gutzwiller and a mean-field approximation, we obtain a Bogoliubov-de Gennes equation
and a set of self-consistent equations in a similar way to the BCS
mean-field theory\cite{hime}:

\begin{equation}
\left(
\begin{array}{cc}
H_{ij} & F_{ij} \\
F_{ji}^{*} & -H_{ji}
\end{array}
\right)
\left(
\begin{array}{c}
u_{j}^{\alpha} \\
v_{j}^{\alpha}
\end{array}
\right)
= E^{\alpha}
\left(
\begin{array}{c}
u_{i}^{\alpha} \\
v_{i}^{\alpha}
\end{array}
\right)
,
\end{equation}
with
\begin{eqnarray}
H_{ij} &=& -\sum_{\tau}\left\{ g_{t}t + \left( \frac{3}{4}g_{s}J
-\frac{1}{4}J_{N}\right) \xi_{ji}
\right\} \delta_{j=i+\tau}   \nonumber \\
& & - \mu\delta_{ij} + V_{i}^{\rm{imp}}\delta_{ij}
\nonumber \\
F_{ij}^{*} &=& -\sum_{\tau}\left( \frac{3}{4}g_{s}J
+ \frac{1}{4}J_{N} \right) \Delta_{ij}\delta_{j=i+\tau}  ,
\end{eqnarray}
where $\tau$ runs as vectors pointing to the nearest-neighbor sites and $g_{t}$, $g_{s}$ are the renormalization factors in the Gutzwiller approximation given as functions of the doping rate $\delta = 1-n$:
\begin{equation}
g_{t} = \frac{2\delta}{1+\delta},~~g _{s} = \frac{4}{(1+\delta)^{2}}.
\end{equation}

Self-consistent equations are
\begin{eqnarray}
\Delta_{ij} &=& \langle c_{i\uparrow}^{\dag}c_{j\downarrow}^{\dag}
\rangle \nonumber \\
&=&-\frac{1}{4}\sum_{\alpha}
(u_{i}^{\alpha *}v_{j}^{\alpha} +
u_{j}^{\alpha}v_{i}^{\alpha *})\tanh\frac{\beta E^{\alpha}}{2} ,
\nonumber \\
\xi_{ij\sigma} &=& \langle c_{i\sigma}^{\dag}c_{j\sigma}
\rangle \nonumber \\
&=& -\frac{1}{4}\sum_{\alpha}
(u_{i}^{\alpha *}u_{j}^{\alpha} -
v_{j}^{\alpha}v_{i}^{\alpha *})\tanh\frac{\beta E^{\alpha}}{2} .
\end{eqnarray}
In the following, we assume $\xi_{ij\uparrow} = \xi_{ij\downarrow}
\equiv \xi_{ij}$ and that $\Delta_{ij}$ is a singlet pairing, i.e.,
$\Delta_{ij} = \Delta_{ji}$.
Since we consider the dilute limit of the impurity concentration, $\mu$ is fixed to the bulk value  $\mu_{0}$ determined without impurities.

\begin{figure}
\vspace{20pt}
\begin{center}
\leavevmode
 \epsfxsize=80mm
 \epsfbox{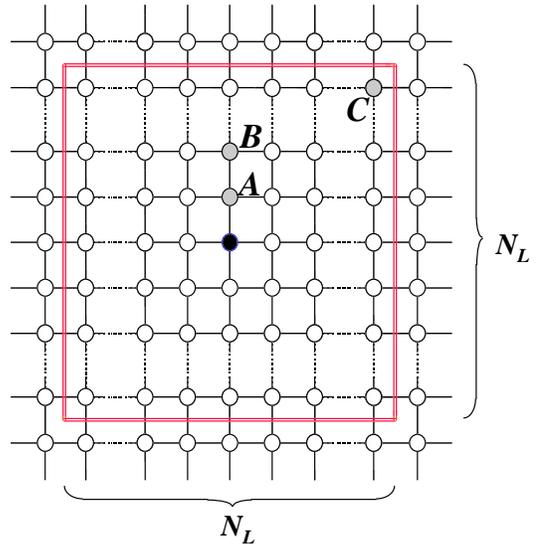}
\end{center}
\caption{
The unit cell for the impurity problem (enclosed by the double lines). 
$N_{L}$ is odd.
The black circle represents the impurity site.
$A$, $B$ and $C$ represent sites where the local DOS is calculated.
}
\label{cell}
\end{figure}

In the present calculation, we regard the $N_{L}\times N_{L}$ square lattice as a unit cell of which the impurity is located at the center ($N_{L}$ is odd. See Fig.~\ref{cell}).
We use periodic boundary conditions in $x$- and $y$-directions and assume a translational symmetry of $\Delta_{ij}$ with respect to this unit cell.
Then we can make use of the Fourier transform of the Bogoliubov-de Gennes equation.
In order to study the effects of a single impurity, it is necessary to take the size of the unit cell large enough.
Specifically, $N_{L}$ should be chosen so that the typical V-shape local density of states is obtained on the sites at the edge of the unit cell.
For the Fourier transform, we take the number of the unit cells $N_{c}=72 ~(8\times 9).$

%

We solve numerically the Bogoliubov-de Gennes equation and carry out an iteration until the self-consistent equations for $\Delta_{ij}$ and $\xi_{ij}$ are satisfied.
We take $J/t = 0.2, J_{N}/t = 0$ and the doping rate $\delta = 0.05$ throughout in this letter.
These parameters are approximately consistent with realistic ones obtained 
by Hybertsen $et ~al$.\cite{hybertsen}
The impurity potential is taken $V_{i}^{\rm{imp}} = 1000t$ on the impurity site so that the impurity scattering is in the unitary limit.

\begin{figure}
\vspace{20pt}
\begin{center}
\leavevmode
 \epsfxsize=80mm
 \epsfbox{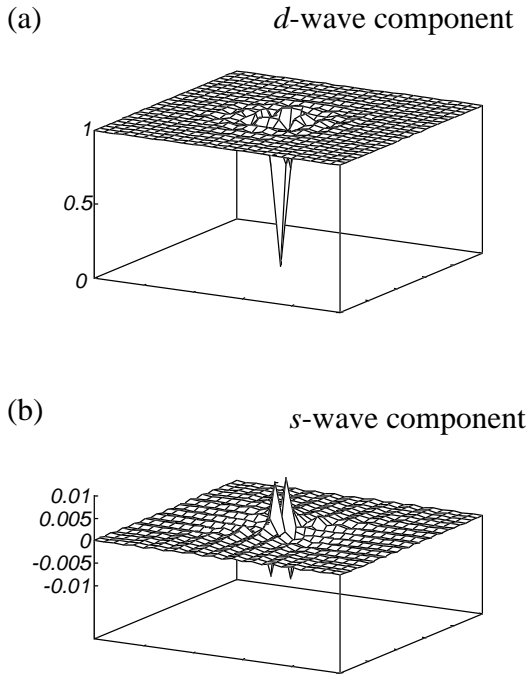}
\end{center}
\caption{Spatial dependence of the relative amplitudes of (a) $d$-wave and (b) $s$-wave components of the order parameters in the unit cell illustrated in Fig.~\ref{cell}.
Note the different scales for (a) and (b).
The order parameters are normalized by the bulk value of the $d$-wave component.
The doping rate is $\delta = 0.05$ and $J/t = 0.2$.}
\label{op}
\end{figure}

Figure~\ref{op} shows the obtained order parameters in the unit cell
for $N_{L} = 27$ and $\delta = 0.05$.
We define $\Delta_{i}^{x}$ and $\Delta_{i}^{y}$ for each site
by averaging $\Delta_{i,i+\tau}$'s on the two opposite bonds of $x$- and $y$-directions.
Furthermore, the obtained $\Delta_{i}^{x}$ and $\Delta_{i}^{y}$ are 
decomposed into extended $s$-wave and $d$-wave components as
\begin{eqnarray}
\Delta_{i}^{x} &=& \Delta_{i}^{d} + \Delta_{i}^{s},
\nonumber \\
\Delta_{i}^{y} &=& -\Delta_{i}^{d} + \Delta_{i}^{s}.
\end{eqnarray}

From Fig.~\ref{op}(a), one can see that the $d$-wave order-parameter is suppressed within an about $5a$ radius around the impurity site, with $a$ being the lattice constant.
Simultaneously, as shown in Fig.~\ref{op}(b), an $s$-wave component is slightly induced with the magnitude of less than  $1\%$ relative to the bulk $d$-wave component.
We note that the calculation with fixed $\xi\equiv\xi_{0}$ ( value for the uniform case ) overestimates the induction of the $s$-wave component up to $8\%$.
Different from the cases with a vortex\cite{hime} and surfaces\cite{tanuma}, the $d$- and $s$-wave components obtained here are real.
The induced $s$-wave component has the following characteristics similar to the results of the previous work\cite{franz} ; (i) it has the same magnitude and opposite sign with $\pi/2$ rotation centered on the impurity site, (ii) it is not induced on the sites in the diagonal direction from the impurity.
These characteristics originate from the sign-changing property of the $d_{x^{2}-y^{2}}$-wave order parameter, and the symmetry around the impurity.

%

\begin{figure}
\vspace{20pt}
\begin{center}
\leavevmode
 \epsfxsize=80mm
 \epsfbox{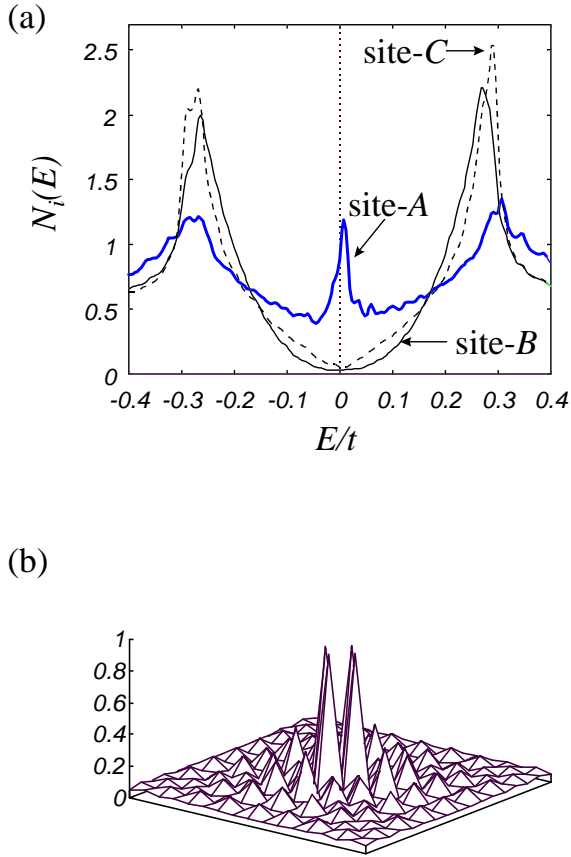}
\end{center}
\caption{
(a) The local density of states at the three sites $A$, $B$ and $C$ (see Fig.1), for $\delta = 0.05$ and $J/t = 0.2$.
The size of the unit cell is $27\times 27$ sites.
$N_{i}(E)$ has a zero-energy peak on the nearest-neighbor sites of 
the impurity site (site-$A$).
We can see that $N_{i}(E)$ reproduces the shape of the uniform $d_{x^{2}-y^{2}}$-wave state on the sites located on the edge of the unit cell (site-$C$). 
(b) Spatial variations of the local density of states at $E=0$ plotted over $17\times 17$ sites around the impurity.
}
\label{ldos1}
\end{figure}

Next, we calculate the local density of states (LDOS) defined as
\begin{eqnarray}
\label{ldos}
N_{i}(E) &=& \frac{1}{N_{c}}\sum_{k,\alpha}\left[ ~|u_{i}^{\alpha}(\mbox{\boldmath $k$})|^{2}~\delta(E^{\alpha}(\mbox{\boldmath $k$})-E) \right. \nonumber \\
&& \left. {}~~~~~~+~|v_{i}^{\alpha}(\mbox{\boldmath$k$})|^{2}~\delta(E^{\alpha}(\mbox{\boldmath $k$})+E) ~\right],
\end{eqnarray}
where $i$ represents a site, {\boldmath$k$} is the Bloch wave number
and $\alpha $ is the eigenstate number.
Though eq.~(\ref{ldos}) is calculated within the Gutzwiller approximation, we can obtain qualitative features\cite{hime}.
Figure~\ref{ldos1} (a) shows the LDOS obtained on the site-$A$, -$B$ and -$C$ illustrated in Fig~\ref{cell}.
Here, we take $N_{c} = 240 ~(15\times 16)$.
On the sites located on the edge of the unit cell (site-$C$), we actually find the typical DOS for $d$-wave superconductivity.
Therefore, the impurity is considered to be well-isolated in the present situation; $N_{L} = 27, ~\delta = 0.05$.
On the nearest-neighbor site of the impurity (site-$A$), the LDOS detects the zero-energy states (ZES) as the zero-energy peak (ZEP). 
This is due to the fact that, in the unitary scattering limit, there always exist some impurity-scattering processes in which the quasiparticles feel the sign-change of the pair potential.
The asymmetric shape of the ZEP is due to the breaking of the electron-hole
symmetry \cite{tsuchi}.

Different from the case with vortex or surface, the splitting of the ZEP does not appear within the present calculation since the induced $s$-wave component is real and rather small.
On the 3rd-neighbor site of the impurity (site-$B$), the ZEP cannot be seen and the weight in the low energy region is reduced compared with the bulk $d$-wave case (site-$C$).
This is due to the spatial variation of the ZES.
In order to see it, the DOS with a fixed energy $E=0$ is plotted as a function of sites around the impurity in Fig.~\ref{ldos1}(b).
This figure represents the spatial oscillating behavior of ZES, which can be regarded as the Friedel oscillation.
This is the reason for the absence of the ZEP on site-$B$.
Furthermore, Fig.~\ref{ldos1}(b) shows that there are no $\frac{1}{r}$-tails along the diagonal direction, and that the ZES is approximately localized around the impurity.
This behavior is consistent with the results in refs.~6 and ~13 but is contrary to the results in refs.~9 and ~11.
Our preliminary results show that whether the ZES localizes or not depends on the interaction strength and the shape of the Fermi surface \cite{tsuchi}.

%
An interesting question is how the ZES behaves in the overdoped region.
To examine the doping dependence of the ZES, attention should be given to the size of the unit cell since the coherence length becomes longer with increasing $\delta$.
Even in the case of $\delta=0.05$, we find that small-size calculations ($N_{L} \leq 23$) give wrong results, i.e., they sometimes give a splitting of the ZEP.
If we increase the size of the unit cell, the width of the splitting shrinks
roughly in proportion to $1/N_{L}$, and the splitting vanishes when $N_{L} \geq
25$.
This size-dependence is due to the intercell overlapping effect of ZES and is not intrinsic to the single impurity case.
Thus the appearance of the splitting of ZEP found by Tanaka $et~al.$\cite{kuboki} is due to the small cluster size they used ($N_{L} = 16\sim18)$.
Here, let us estimate the size of the unit cell required in the case of $\delta = 0.1$.
The obtained energy gap without any impurity is $\Delta = 0.27t$ at $\delta = 0.05$ and $\Delta = 0.22t$ at $\delta = 0.1$.
The Fermi energy $\epsilon_{F}$ is estimated from $\epsilon_{F} \sim 4(g_{t}t + \frac{3}{4}g_{s}J|\xi_{0}|)$.
We obtain $\epsilon_{F}\sim 0.79t$ at $\delta = 0.05$ and $\epsilon_{F}\sim 1.11t$ at $\delta = 0.1$.
Then the ratio of the coherence length at $\delta = 0.05$ to $\delta = 0.1$ is found to be $(\frac{\sqrt{\epsilon_{F}}}{\Delta} |_{\delta = 0.1}) / (\frac{\sqrt{\epsilon_{F}}}{\Delta} |_{\delta = 0.05}) \sim 1.5$.
Thus, we can estimate the size of the unit cell required for $\delta = 0.1$ as $N_{L} \sim 37$.
Indeed, our preliminary calculation with $N_{L} = 37$ shows an almost single peak near the zero-energy in the LDOS.
Although this suggests the similar results in $\delta = 0.1$, the doping dependence of the ZES from under- to overdoped region is to be published elsewhere\cite{tsuchi}.

In this letter, we have shown for the first time that the ZES gives ZEP in the LDOS around a nonmagnetic impurity in the $d_{x^{2}-y^{2}}$-wave superconducting state of the underdoped $t$-$J$ model.
In the scanning tunneling spectroscopy (STS) experiment where the direct observation of the LDOS is possible with atomic-scale spatial resolution, the ZBCP, which corresponds to the ZEP obtained here, is expected to be probed just as in the case of a well-oriented (110) surface\cite{alff}.
Furthermore we found that the ZES is approximately localized (radius 5$a$) around the impurity.
The induced $s$-wave component is real but small and thus there is no splitting of the ZEP. 

H. T. and Y. T. wish to thank J. Inoue and H. Ito for their useful discussions.
H. T. also thanks Y. Nakamoto for useful advice on numerical works.
This work is supported in part by a Grant-in-Aid for Creative Basic Research (08NP1201) of the Ministry from Education, Science, Sports and Culture.
Numerical computation in this work was partially carried out at the Yukawa Institute Computer Facility and the Supercomputer Center, Institute for Solid State
Physics, University of Tokyo.

\end{document}